\documentclass{ws-fml}
\usepackage{multicol}
\begin{document}
\bibliographystyle{ws-fml}

\catchline{1}{1}{2008}{}{}
\markboth{David J. Singh}{Thermopower of SnTe from Boltzmann Transport
Calculations}

\title{Thermopower of SnTe from Boltzmann Transport Calculations}

\author{David J. Singh}

\address{Materials Science and Technology Division,
Oak Ridge National Laboratory\\
Oak Ridge, TN 37831-6114, USA.
\email{singhdj@ornl.gov} }

\maketitle

\begin{history}
\received{Day Month Year}
\revised{Day Month Year}
\end{history}

\begin{abstract}
The doping and temperature dependent thermopower of SnTe
is calculated from the first principles band structure using
Boltzmann transport theory. We find that the $p$-type
thermopower is inferior to PbTe consistent with experimental observations,
but that the $n$-type thermopower is substantially more favorable.
\end{abstract}


\begin{multicols}{2}

\section{Introduction}
Thermoelectrics offer a useful technology for small scale cooling
and power generation applications. The performance of such devices
is limited by a materials and temperature dependent dimensionless figure
of merit, $ZT=\sigma S^2 T/\kappa$,
where $T$ is temperature, $\sigma$ is the electrical conductivity,
$S$ is the thermopower and $\kappa$ is the thermal
conductivity, which is generally
the sum of lattice and electronic components,
$\kappa_l$ and $\kappa_e$.
There has been considerable recent interest in finding new thermoelectric
materials.\cite{snyder,sootsman}

In this regard, there have been recent developments in PbTe based materials,
particularly with heavy Tl doping \cite{heremans} and via alloying
with small amounts of
AgSbTe$_2$, which leads to a nanostructured material denoted LAST.
\cite{hsu}
Both of these materials have $ZT$ well above unity.
In the LAST material this arises primarily because of a reduction
in thermal conductivity associated with the nanostructuring,
\cite{hsu}
and possibly connected with the nearness of the phase to a ferroelectric
instability. \cite{an}
In PbTe:Tl the high $ZT$ is instead caused by an enhanced thermopower
at high Tl doping levels.
GeTe alloyed with small amounts of AgSbTe$_2$ is also a very high
performance thermoelectric, with a $ZT$ that exceeds 1.2
at high temperature. \cite{snyder,cook}
There are, however, certain disadvantages to these materials. Specifically,
the use of Pb limits applications due to environmental and regulatory issues,
while Ge is a very costly material. On the other hand, Sn is both
environmentally benign and inexpensive.
This suggests investigation of SnTe, which occurs in the same
structure.

In fact, there have been a number of experimental investigations of
the thermoelectric performance of SnTe with various doping strategies.
One difficulty is that SnTe forms with a high concentration
of native defects, particularly Sn vacancies, which are difficult to 
avoid because of the shape of the liquidus line in the Sn-Te phase
diagram. \cite{breb1}
This leads to difficult to control high $p$-type carrier concentrations.
These can be compensated by alloying with Bi. However, even then samples are
still generally $p$-type.
\cite{brebrick,tamor}

Furthermore, experimental
investigations of $p$-type SnTe have invariably found low
values of $ZT$ in comparison to PbTe. \cite{wu}
This can be traced to the behavior of $S(T)$. The thermal conductivity
of SnTe, $\kappa\sim$ 2 W/m K, is similar to that of GeTe and PbTe,
and can be strongly reduced by alloying. \cite{wu} However,
$S(T)$ is lower than in comparable PbTe. \cite{brebrick}
This leads to a much lower $ZT$ in SnTe as compared with PbTe.
Interestingly, an enhancement of the thermopower was found
at high carrier concentrations between $\sim 2$x10$^{20}$ cm$^{-3}$
and 10$^{21}$ cm$^{-3}$, \cite{brebrick}
similar to what occurs in Tl doped PbTe, although in PbTe this is at
lower carrier density. \cite{singh-pbte}

\section{Structure and Methods}

We applied Boltzmann transport theory within the constant scattering
time approximation (CSTA) to the first principles electronic structure as
obtained within density functional theory using the general potential
linearized augmented planewave (LAPW) method
as implemented in the WIEN2K code, \cite{wien}
similar to our prior work on
PbTe and PbBi$_2$Te$_4$. \cite{singh-pbte,zhang}
Spin-orbit was included in all the calculations.
We used well converged basis sets and Brillouin zone samplings with
LAPW sphere radii of 2.0 Bohr for both Sn and Te.
The CSTA allows the thermopower to be directly calculated from the
band structure as a function of carrier concentration and $T$ with
no adjustable parameters. \cite{mazin,ong}
The needed integrals were obtained using the
BoltzTraP code. \cite{boltztrap} One complication is that the band
gaps of semiconductors are typically underestimated in density functional
calculations. For small band gap materials this will lead to bipolar
conduction and reduce the thermopower at high temperatures especially
when the carrier density is low. We used the Engel-Vosko generalized
gradient approximation (GGA) to avoid this. \cite{ev}
Unlike standard GGA's,
this functional is optimized to reproduce the exchange-correlation
potential rather than the total energy, and as a result gives improved
band gaps.

While PbTe is almost ferroelectric, both SnTe and GeTe are ferroelectric,
with Curie temperatures, $T_C$ of 120 K and $\sim 700$ K, respectively.
SnTe occurs in the cubic NaCl structure above $T_C$ and is rhombohedral
below. Here we present results from 300 K up, and so we use the NaCl
structure with the experimental lattice parameter, $a$=6.303 \AA.

\section{Electronic Structure}

The calculated band structure is shown in Fig. \ref{bands}.
It is similar to prior relativistic calculations
\cite{tung}
and in particular
shows a direct gap at the $L$ point, with non-parabolic bands that
become effectively heavier as the chemical potential moves away from
the band edges.
Our calculated band gap with the Engel-Vosko GGA is $E_g$=0.17 eV,
which is less than the value
of 0.30 eV obtained by the same procedure applied to PbTe. \cite{singh-pbte}
The calculated gap is in good agreement with the experimental
room temperature value of 0.18 eV
\cite{rogers}
(note that at low $T$ SnTe has a different ferroelectric crystal
structure).

\begin{figurehere}
\centerline{\psfig{file=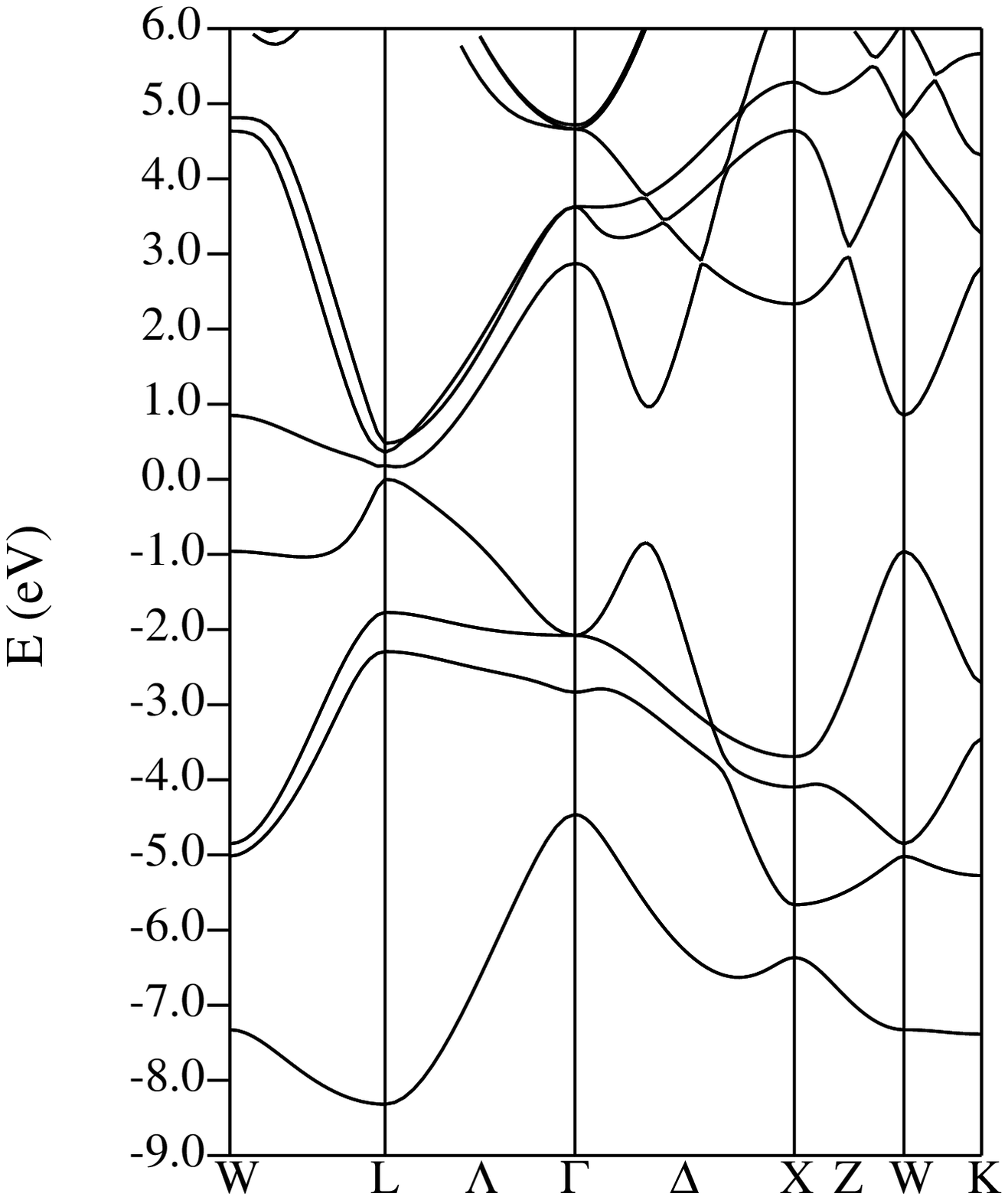,width=\columnwidth}}
\caption{Band structure of cubic SnTe obtained with the Engel-Vosko
GGA, including spin-orbit.}
\label{bands}
\end{figurehere}

\begin{figurehere}
\centerline{\psfig{file=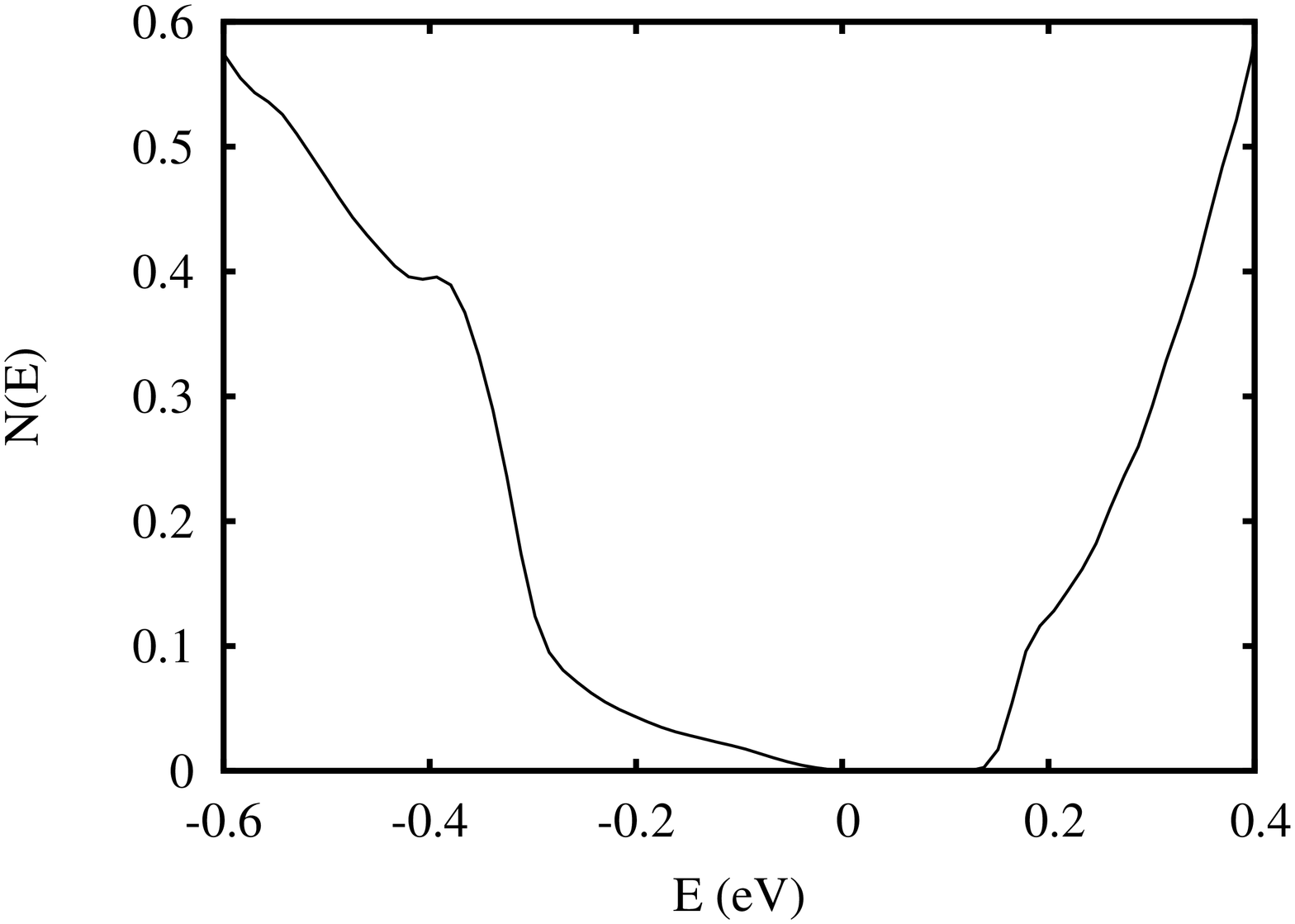,angle=270,width=0.95\columnwidth}}
\caption{Electronic DOS of SnTe.}
\label{dos}
\end{figurehere}

The calculated electronic density of states (DOS) in the region near
the band edges is shown in Fig. \ref{dos}. Similar to PbTe, \cite{singh-pbte}
it shows a light mass behavior at the valance band edge with a strong upturn
at $\sim$ 0.3 eV. This arises for the same reason, i.e. connection of the
$L$-point pockets along the approximately [001] oriented directions connecting
them. This is qualitatively consistent with the band structure model
developed by Allgaier and Houston based on magnetotransport measurements.
\cite{allgaier}
However, this upturn in the DOS is not as strong as in PbTe.
Also, one may note in the band structure that the dispersion of the conduction
bands from $L$-$W$ is weaker in SnTe. This leads to a higher DOS near the
conduction band edge, i.e. higher effective band mass
for $n$-type. Finally, the DOS
in the conduction bands has structure at $\sim0.2$ eV above the edge.
This comes from the second and third conduction bands which have their minima
at $L$ (note that there is a small splitting between these minima).
These bands are more dispersive than the lowest conduction band.

Thus, SnTe behaves as a material with lighter bands than PbTe for $p$-type
and heavier bands than PbTe for $n$-type.
These differences
are reflected in the behavior of the thermopower.

\section{Thermopower}

Our main results are presented in Figs. \ref{doping-p} and \ref{doping-n},
which show the doping dependence of $S(T)$ for $p$-type
and $n$-type, respectively. We show data from 300 K to 800 K
(the liquidus temperature is 1078 K \cite{breb1}).
As may be seen, the $p$-type thermopowers are inferior to PbTe.

It can be shown that when the Wiedemann-Franz relation holds
$ZT=rS^2/L$, where $L$ is the Lorentz number and 
$r=\kappa_e/\kappa < 1$. Taking the standard value of $L$, $S >$157 $\mu$V/K
is required for $ZT$=1, even assuming that $\kappa_l$ is negligible.
Based on this and the results in Fig. \ref{doping-p}
it would seem that the only regimes where $p$-type SnTe could have
a high $ZT$ are (1) at high carrier concentrations ($\sim10^{20}$ cm$^{-3}$)
and very high temperatures, where the material is known to be prone to
Sn vacancy formation and
(2) near ambient $T$ with a low carrier concentration
$\sim$ 2x10$^{18}$, and then only if very high mobility samples
can be prepared or the thermal conductivity suppressed (e.g. by alloying).
Interestingly, we do find an enhancement of the thermopower in the range
from 10$^{20}$ cm$^{-3}$
to 10$^{21}$ cm$^{-3}$ as in experiment.
However, unlike PbTe \cite{singh-pbte}, the resulting values of $S$ are too
low for high $ZT$.

The situation for $n$-type is much more interesting.
At very low $n$-type carrier densities ($\sim10^{18}$ cm$^{-3}$)
and very high temperatures, the
combination of light valence bands and heavy conduction bands
leads to a positive $T$ independent thermopower due to bipolar conduction.
At more realistic doping levels the thermopower is much higher than
for $p$-type, and importantly $S(T)$ is also substantially higher than in
$n$-type PbTe and remains high even at very high doping levels in excess of
10$^{20}$ cm$^{-3}$, depending on $T$.

\begin{figurehere}
\centerline{\psfig{file=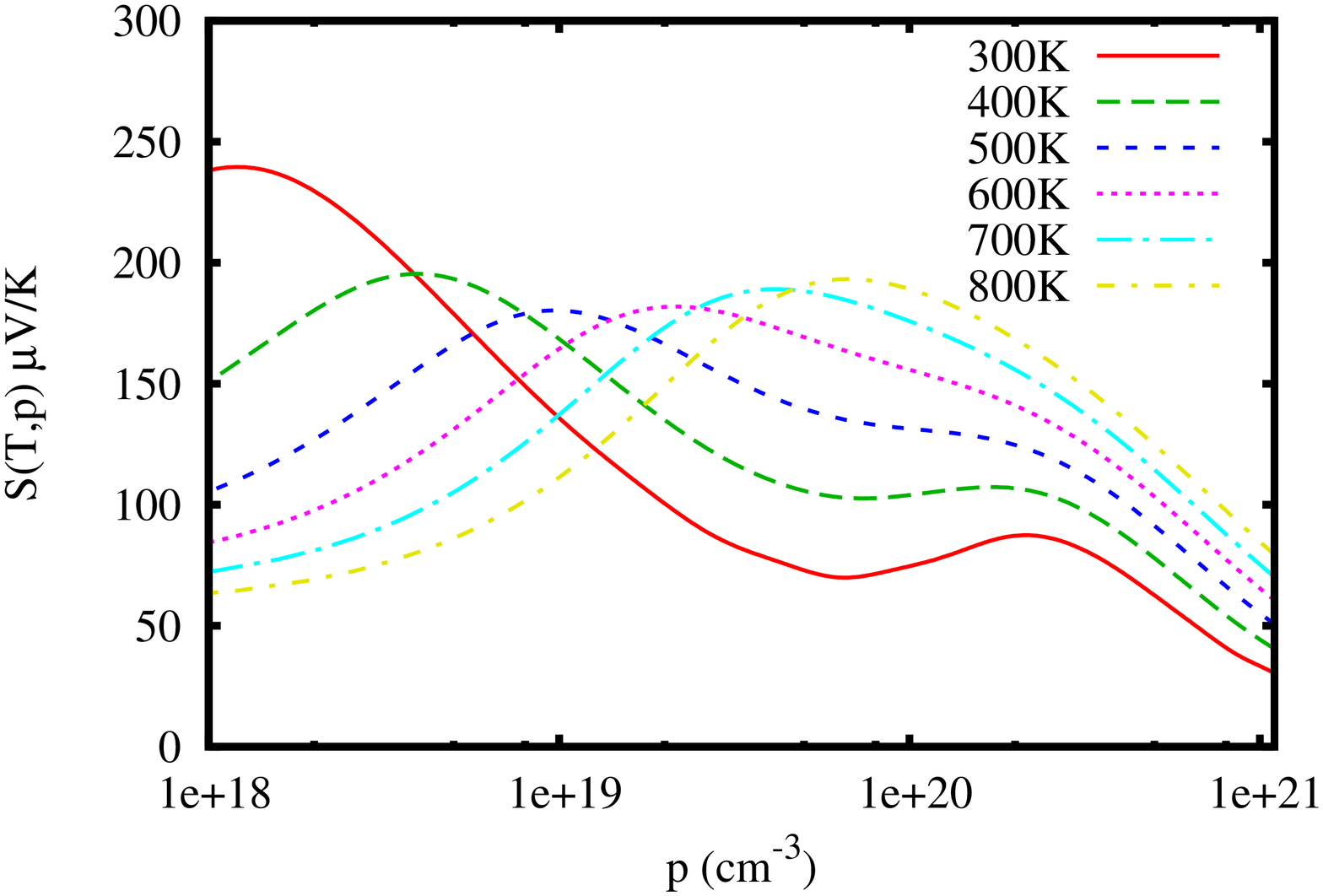,angle=270,width=\columnwidth}}
\caption{Calculated doping dependence of $S(T)$ for $p$-type SnTe.}
\label{doping-p}
\end{figurehere}

\begin{figurehere}
\centerline{\psfig{file=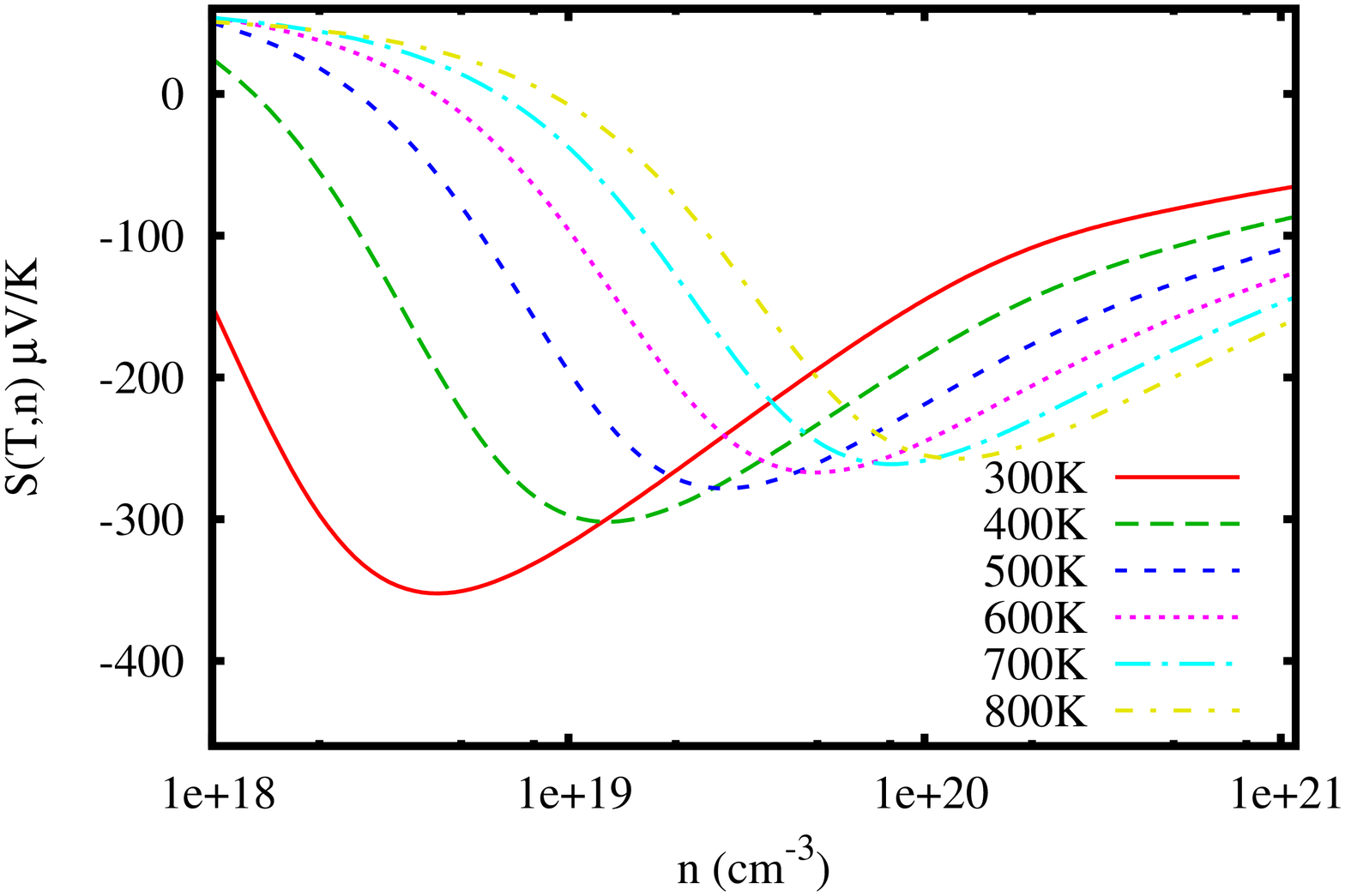,angle=270,width=\columnwidth}}
\caption{Calculated doping dependence of $S(T)$ for $n$-type SnTe.}
\label{doping-n}
\end{figurehere}

\section{Summary and Conclusions}

To summarize, we calculated $S(T)$ for SnTe as a function of
carrier concentration and temperature.
We find that the values of $S$ are substantially inferior to those
of PbTe for $p$-type, consistent with existing experimental data
showing inferior thermoelectric performance.
On the other hand, we find that the behavior of $S$ for $n$-type is
significantly superior to that of $n$-type PbTe. Considering that
the two compounds have similar lattice thermal conductivities
this indicates that optimized
$n$-type SnTe could have superior thermoelectric
performance to $n$-type PbTe.
However, as mentioned, SnTe almost invariably forms as $p$-type
material due to the shape of the liquidus line in the Sn-Te binary phase
diagram, which results in Sn deficient samples. This can be
controlled in part by annealing or Bi doping. \cite{breb1,brebrick}
However, there has been little recent experimental work on SnTe,
perhaps because of the known poor performance of $p$-type material.
Perhaps modern doping strategies, such as controlled co-doping
(possibly including the Te site, e.g. alloying with both Bi and Se)
could yield $n$-type material. If so, based on the present results
it will be very interesting
to determine the thermoelectric performance of heavily doped
$n$-type SnTe.

\nonumsection{Acknowledgments} 
\noindent We are grateful for helpful discussions with David Parker.
This research was sponsored by the U.S. Department of Energy,
Office of Science, Materials Sciences and Engineering Division.

\bibliography{SnTe}

\end{multicols}
\end{document}